\documentclass[conference]{IEEEtran}
\usepackage{cite}
\usepackage{amsmath,amssymb,amsfonts}
\usepackage{algorithmic}
\usepackage{graphicx}
\usepackage[T1]{fontenc}
\usepackage[utf8]{inputenc}
\usepackage{textcomp}
\usepackage{tabularx}
\usepackage{comment}
\usepackage{braket}
\usepackage{threeparttable}
\usepackage{booktabs}
\usepackage{pifont}
\usepackage{subcaption}

\usepackage{url}
\usepackage{multirow}
\usepackage{makecell}
\usepackage{dblfloatfix}

\begin{document}

\title{Low-Latency FPGA Control System for Real-Time Neural Network Processing in CCD-Based Trapped-Ion Qubit Measurement}

\author{
    \IEEEauthorblockN{
        Binglei~Lou\textsuperscript{1},
        Gautham~Duddi~Krishnaswaroop\textsuperscript{1}\textsuperscript{†},
        Filip~Wojcicki\textsuperscript{2}\textsuperscript{†},
        Ruilin~Wu\textsuperscript{1},
        Richard~Rademacher\textsuperscript{1},\\
        Zhiqiang~Que\textsuperscript{2},
        Wayne~Luk\textsuperscript{2},
        and~Philip~H.W.~Leong\textsuperscript{1}
    }
    \IEEEauthorblockA{
        \textsuperscript{1}The University of Sydney, Australia;
        \textsuperscript{2}Imperial College London, UK; {\textsuperscript{†} Equal Contribution}\\
        Emails: \{binglei.lou, ruilin.wu, richard.rademacher, philip.leong\}@sydney.edu.au; gdud9079@uni.sydney.edu.au;\\
        \{faw18, z.que, w.luk\}@imperial.ac.uk
    }
}

\maketitle

\begin{abstract}
Accurate and low-latency qubit state measurement is critical for trapped-ion quantum computing. While deep neural networks (DNNs) have been integrated to enhance detection fidelity, their latency performance on specific hardware platforms remains underexplored. This work benchmarks the latency of DNN-based qubit detection on field-programmable gate arrays (FPGAs) and graphics processing units (GPUs). The FPGA solution directly interfaces an electron-multiplying charge-coupled device (EMCCD) with the subsequent data processing logic, eliminating buffering and interface overheads. As a baseline, the GPU-based system employs a high-speed PCIe image grabber for image input and I/O card for state output. We deploy Multilayer Perceptron (MLP) and Vision Transformer (ViT) models on hardware to evaluate measurement performance. Compared to conventional thresholding, DNNs reduce the mean measurement fidelity (MMF) error by factors of 1.8-2.5$\times$ (one-qubit case) and 4.2-7.6$\times$ (three-qubit case). FPGA-based MLP and ViT achieve nanosecond- and microsecond-scale inference latencies, while the complete single-shot measurement process achieves over 100× speedup compared to the GPU implementation. Additionally, clock-cycle-level signal analysis reveals inefficiencies in EMCCD data transmission via Cameralink, suggesting that optimizing this interface could further leverage the advantages of ultra-low-latency DNN inference, guiding the development of next-generation qubit detection systems.
\end{abstract}

\section{Introduction}
\label{sec:introduction}

\begin{figure*}[b]
\centerline{\includegraphics[width=0.86\linewidth]{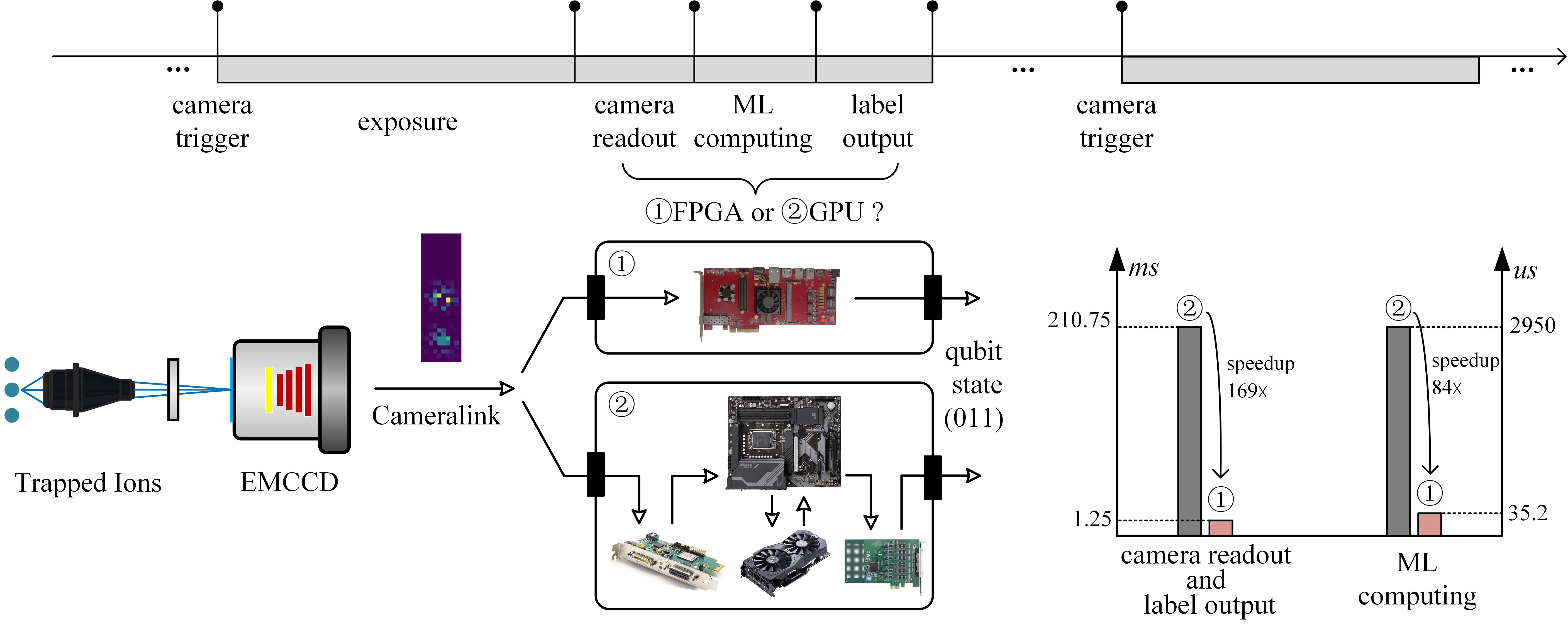}}
\caption{In the block diagram of ML-aided qubit detection in a trapped ion system, photons scattered by the ions are collected and magnified by optical elements before being directed towards an EMCCD camera. The resulting images are transmitted to the acceleration hardware, such as an FPGA or GPU, via the Cameralink protocol. The right bar chart illustrates the speedup achieved by the DNN-based (ViT) measurement solution on a 3-qubit test.}
\label{fg:intro_figure}
\end{figure*}

Quantum algorithms have the potential for new and highly efficient processing of large data, including super-polynomial speedup of search~\cite{grover2005fixed}, cryptography~\cite{shor1994algorithms}, and random walk~\cite{childs2003exponential} problems, along with many others~\cite{montanaro2016quantum}. However, the implementation of the apparatus itself is challenging~\cite{wineland1998experimental}. Trapped ion quantum computers are one of several promising platforms~\cite{trapion1998,trapion2006}, where, in a typical blade-style Paul trap, individual atoms are held in space by radio frequency (RF) electric fields~\cite{milne2021phd}. When ionized, they behave like hydrogenic atoms, with energy levels like those for Ytterbium shown in Fig.~\ref{fg:171Yb}. The qubit state of each atom is the electronic sublevel of the valence electron. Tight confinement of ions in the chain also experiences Coulomb repulsion, and exchange motional energy to form the multi-qubit operations required to produce entangled states~\cite{Quantumentanglement}.

Qubit state measurement in trapped-ion systems is an integral part of quantum information processing. This can be performed via a state-dependent fluorescence technique~\cite{trapion1998,trapion2006}. A laser excites a transition, inducing photon scattering if the qubit is in the $\ket{1}$ state, while no photons are scattered in the $\ket{0}$ state due to off-resonant excitation. The detected photon count, captured by an electron-multiplying charge-coupled device (EMCCD) camera, provides a way to determine the qubit state while also enabling spatial resolution of multiple trapped ions~\cite{manovitz2016, schwegler2018_masterthesis, edmunds2021scalable}.

Given that quantum systems are highly susceptible to errors from decoherence~\cite{Decoherence,decoherence_QEC}, error correction must detect and correct these errors before they propagate and accumulate beyond a tolerable threshold. This is essential for achieving fault-tolerant quantum computing and involves measuring error syndromes followed by conditional corrective operations~\cite{QEC,Wang2011}. For these strategies to be effective, qubit state measurements must be both accurate, to correctly identify errors, and fast enough to outpace qubit decoherence. Additionally, minimizing classification latency is crucial to prevent it from becoming a bottleneck that limits the quantum computer’s overall clock speed. These requirements present two key challenges for qubit measurement in single-shot operations: high fidelity and low latency.

Machine learning (ML) techniques have demonstrated high-fidelity qubit detection using images captured by highly sensitive cameras~\cite{jeong2023using,khomchenko2024detection}. Jeong et al.~\cite{jeong2023using} employed a convolutional neural network (CNN) to enhance four-qubit state measurements using an EMCCD camera, achieving a fidelity error reduction of approximately 50\% compared to the maximum likelihood method~\cite{maximum_likelihood}. However, their analysis was performed offline without considering latency. Khomchenko et al.~\cite{khomchenko2024detection} benchmarked various ML methods, including $k$-means clustering, support vector machines (SVMs), and CNNs, for detecting quantum states of ytterbium ions in a Paul trap. While these methods improved fidelity over traditional thresholding, they were deployed on a central processing unit (CPU), resulting in computation latencies ranging from 35$\sim$1930 $ms$.
 
Despite the superior fidelity of existing ML-based qubit detection solutions~\cite{jeong2023using, khomchenko2024detection}, their high computational latency limits their applicability in real-time quantum error correction, highlighting a significant gap in the literature. Furthermore, although field-programmable gate arrays (FPGAs) have been widely employed in trapped-ion systems for low-latency processing, their primary use has been in controller roles, without fully leveraging the advantages of ML representations~\cite{artiq, zhu2023fpga}.

\begin{figure*}[!b]
    \centering
    \includegraphics[width=0.8\linewidth]{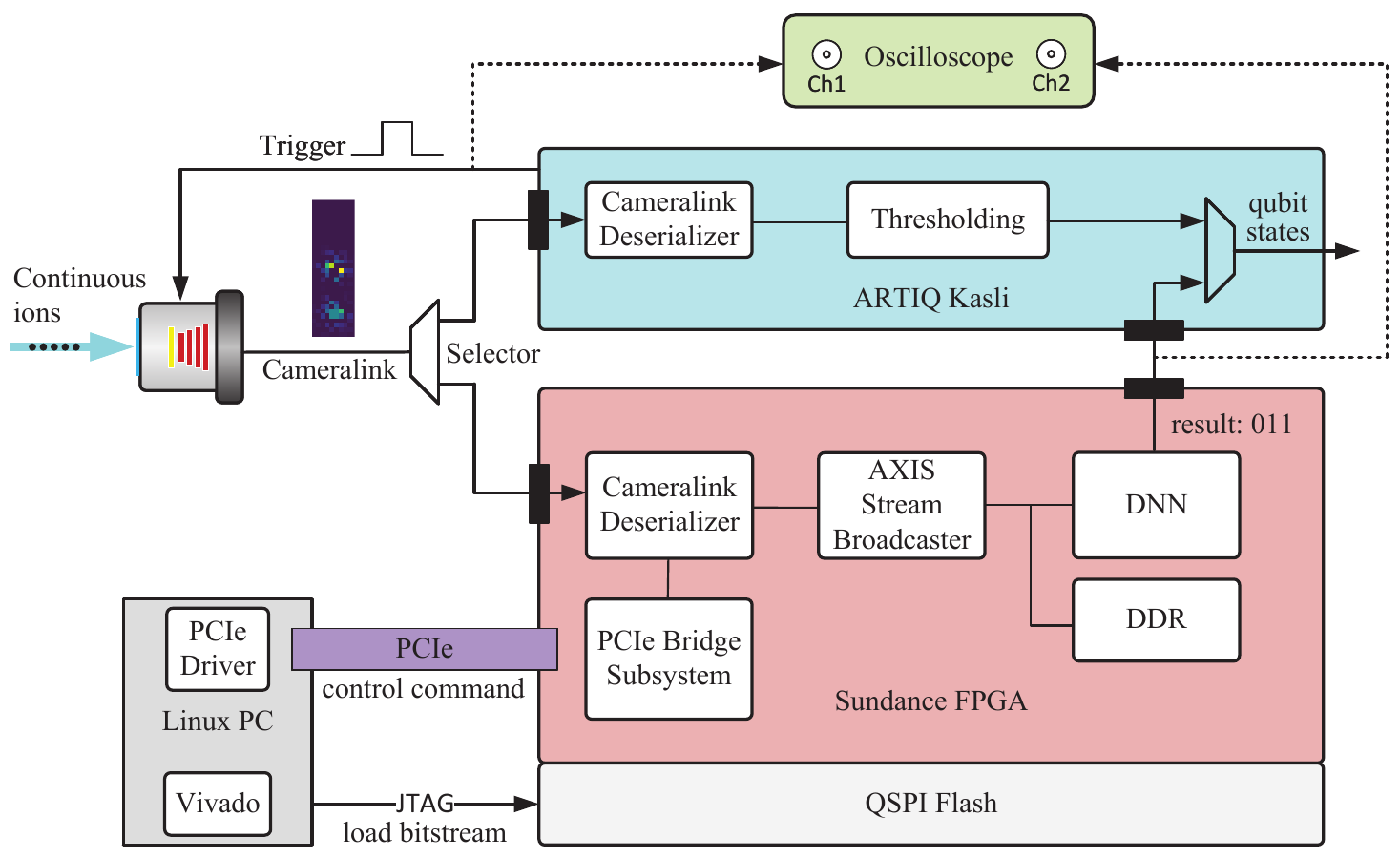}
    \caption{This figure illustrates the framework of our qubit detection system. The camera is triggered by an ARTIQ Kasli system~\cite{artiq,Kasli}, and its images can be processed via either Kasli-based thresholding or FPGA-based DNN processing through the Cameralink protocol.}
    \label{fg:framework}
\end{figure*}

This motivates us to develop a low-latency ML-based qubit detection system using high-resolution camera images and benchmark its performance on modern ML acceleration platforms, including FPGAs and graphics processing units (GPUs).

Our study implements two deep neural network (DNN) models---Multilayer Perceptron (MLP) and Vision Transformer (ViT)---for fast ML-based qubit detection. MLP serves as a universal approximator, capable of representing any continuous function~\cite{hornik1989multilayer}, while ViT is an advanced image classifier that achieves state-of-the-art performance on various image recognition benchmarks~\cite{vit,vit1616}.

For hardware deployment, we primarily focus on FPGA implementation while also developing a GPU-based solution as a baseline. As discussed in the results section, our findings indicate that FPGAs are more suitable for real-time qubit detection. Specifically, we address latency on FPGAs by interfacing the EMCCD camera directly with the DNN detector on the FPGA, eliminating buffering and interface overheads. For DNN computation, we applied a novel architectural implementation of the MLP using a lookup table (LUT)-based technique, achieving nanosecond-level inference latency. The ViT model is deployed using high-level synthesis (HLS), enabling a more balanced trade-off between latency (microsecond-level) and accuracy.

In contrast, our GPU implementation interfaces with a host desktop computer via a high-speed Peripheral Component Interconnect Express (PCIe) rooted image grabber card and an Input/Output (I/O) card, serving as a baseline for performance comparison.

Fig.~\ref{fg:intro_figure} shows the overview of this work with \ding{172} representing the proposed FPGA qubit detection system and \ding{173} as its functionality-equivalent GPU baseline. The ${}^{171}{\text{Y}}{{\text{b}}^ + }$ ions are used in our trapped ion experiment, with their scattered photons collected by an EMCCD camera. Given that the typical EMCCD cameras (e.g., Andor iXon 897~\cite{andor987}) output an image through the Cameralink protocol~\cite{cameralink} and act as a serializer, the FPGA logic detailed in Fig.~\ref{fg:framework} serves as an integrated qubit detection hardware, encapsulating a well-defined workflow to: (1) deserialize the image from cameralink, (2) process the image classification in a DNN accelerator, (3) output results (qubit states) to the next stage. Specifically, (1) and (2) are operated in a pipelined manner through an AXI-Stream (AXIS) bus~\cite{axis} for efficient data processing. A handshakeless Input/Output interface is employed in step (3) to minimize output latency.

\section{Design Implementation}

\subsection{FPGA-based Control System}

\begin{figure*}[]
\centerline{\includegraphics[width=0.8\linewidth]{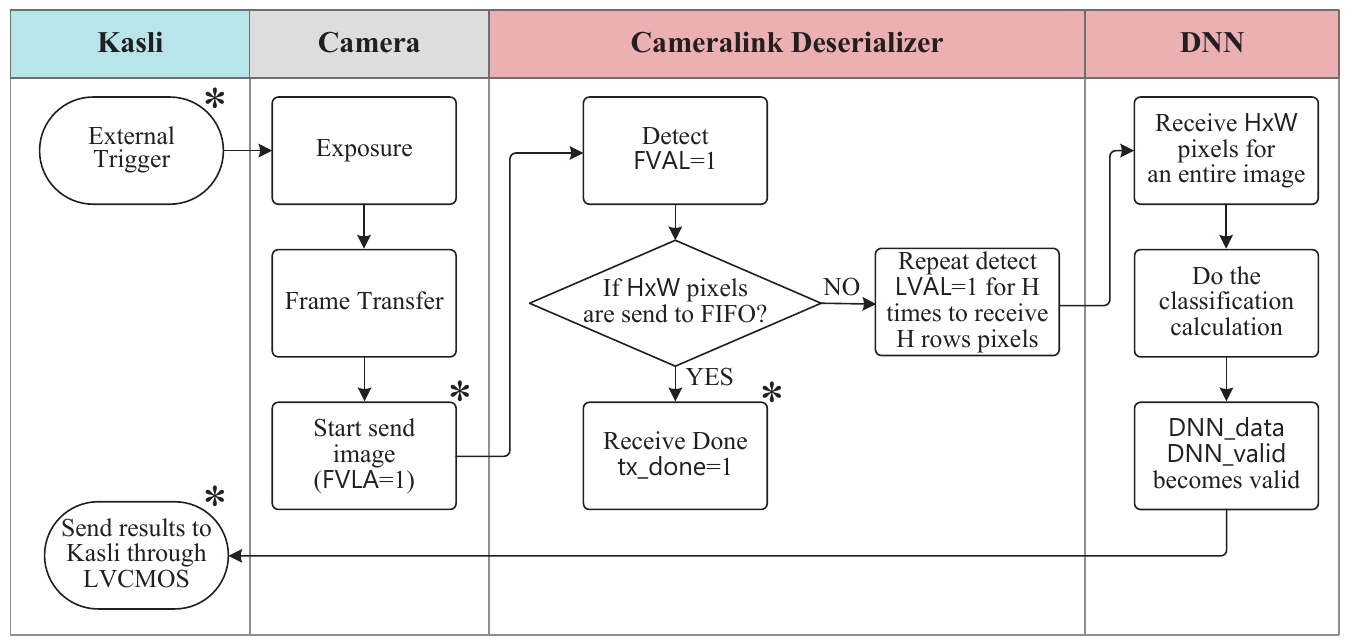}}
\caption{This figure describes the flow chart of the proposed FPGA-based qubit detection system. The steps with the attached $\ast$ symbolize that probe signals are available in this step for latency measurements. The detailed timing diagram of \texttt{FVAL}, \texttt{LVAL}, \texttt{tx\_done} and \texttt{DNN\_valid}, \texttt{DNN\_data} are illustraed in Fig.~\ref{fg:cameralink_timing}.}
\label{fg:flowchart}
\end{figure*}

As depicted in Fig.\ref{fg:framework}, the ion-trap experiment and camera are controlled by the Advanced Real-Time Infrastructure for Quantum Physics (ARTIQ) system\cite{artiq}. ARTIQ's master controller, Kasli~\cite{Kasli}, manages sub-modules such as a Digital IO (DIO) card for camera triggering and a frame-grabber card for image data transfer. A thresholding method~\cite{halama2022real} is implemented on Kasli as a conventional baseline for qubit detection (using its CPU architecture). An auxiliary Sundance FPGA board~\cite{sundance} is employed as our DNN inference platform.

Before each experiment, the EMCCD camera captures trapped ion fluorescence, which is transferred to Kasli/FPGA through the low-latency Cameralink protocol~\cite{cameralink}. Inside the FPGA, a deserializer block decodes the low-voltage differential signaling (LVDS) inputs of the Cameralink interface. The deserialized data is then split into two parallel dataflows via an AXIS Broadcaster~\cite{axis}: one stream buffers the captured images in Double Data Rate (DDR) memory, while the other sends them to the DNN for real-time classification. The classification results are directly transferred to ARTIQ Kasli via a handshakeless IO interface (Fig.~\ref{fg:cameralink_timing}) for minimal latency.

\subsection{Data Flow and Timing Measurements}

\begin{figure}
\centerline{\includegraphics[width=0.99\linewidth]{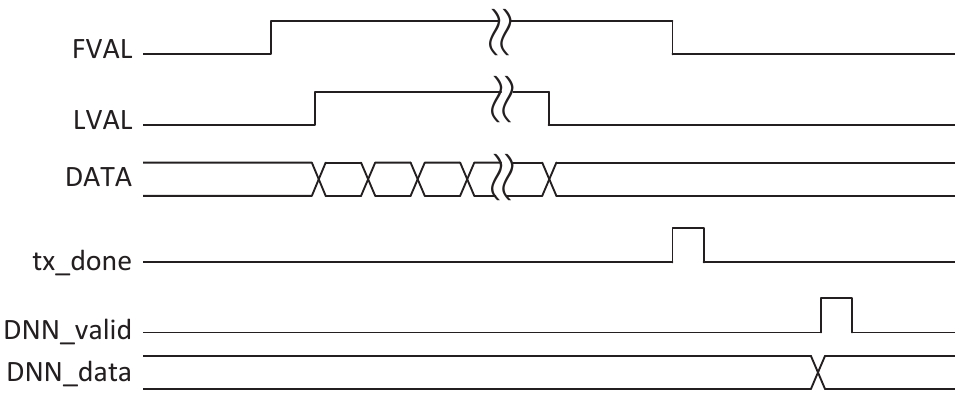}}
\caption{The focus here is mainly on the relationships between \texttt{FVAL}, \texttt{tx\_done}, and the \texttt{DNN} output, with other signal details being omitted for simplicity (We refer to Ref.~\cite{cameralink} for a specialized description of the Cameralink Protocol).}
\label{fg:cameralink_timing}
\end{figure}

Following the framework in Fig.\ref{fg:framework}, the flowchart of the proposed FPGA-based qubit detection system is shown in Fig.\ref{fg:flowchart}. Kasli triggers the camera externally, and the `Frame Transfer' block converts the captured analog image into a digital format. Each digital 2D image, defined as ${(H \times W)}$ pixels, is streamed to the FPGA via the Cameralink protocol. The FPGA's deserializer receives the pixel stream sequentially and buffers it using a double-buffered FIFO to ensure continuous data flow.

As the final pixel of an image is written into the FIFO's read buffer, the preceding ${(H \times W - 1)}$ pixels are already transferred into the DNN accelerator’s internal cache, eliminating data availability delays. Once the full image is assembled, inference begins. Upon completion, the result is stored in \texttt{DNN\_data}, and \texttt{DNN\_valid} is set to \texttt{HIGH}, interfacing with the Kasli controller.

Fig.~\ref{fg:cameralink_timing} illustrates the timing diagram of critical measurement signals. Frame Valid (\texttt{FVAL}) and Line Valid (\texttt{LVAL}) signals indicate a valid frame and line, respectively, and, along with (image) \texttt{DATA}, are provided by the EMCCD camera, which serves as the master. The first rising edge of \texttt{FVAL} marks the start of image transmission to the FPGA. The \texttt{tx\_done} signal, generated within the Cameralink deserializer, is set to \texttt{HIGH} when the received pixel count matches the predefined image resolution, indicating that the entire image has been received and stored in the FIFO.

To evaluate system latency, \texttt{FVAL}, \texttt{tx\_done}, and \texttt{DNN\_valid} are also routed to the FPGA’s output IO as debug signals, allowing precise timing measurements using an external oscilloscope. Steps that can be directly probed are marked with a $\ast$ in Fig.~\ref{fg:flowchart}.

\subsection{MLP Accelerator}
\label{se:MLP Accelerator}

When MLP is applied as the DNN accelerator, the 2D input image is first reshaped into a sequence of flattened 1D vectors.
Inside the MLP, each neuron's output is given by 
\begin{equation}
y = \sigma \left( {\sum\nolimits_{k = 0}^{N-1} {{w_k}{x_k} + b} } \right)
\label{eq:neuron}
\end{equation}
where $N$ is the number of inputs, $x_k$ are the inputs, $w_k$ are the weights, $b$ is a bias term and $\sigma(\cdot)$ is the activation function. 

Given that the state-of-the-art MLP implementation with the lowest latency in the literature is lookup table (LUT)-based DNNs. We deployed the LUT-MLP to benchmark our ultra-low latency qubit detection solution. This is done by combining the multiplication, sum, and activation operations of Eq.~\eqref{eq:neuron} in a single LUT.

\begin{figure}[b]
\centerline{\includegraphics[width=1.0\linewidth]{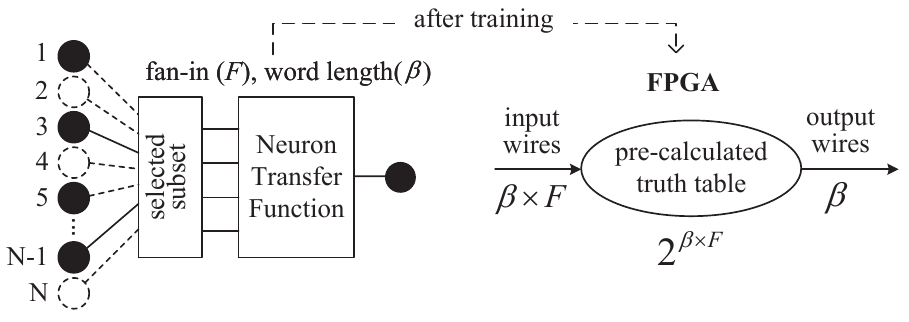}}
\caption{LUT-DNN Method.}
\label{fg:lut1}
\end{figure}

\begin{figure}
\centerline{\includegraphics[width=1.0\linewidth]{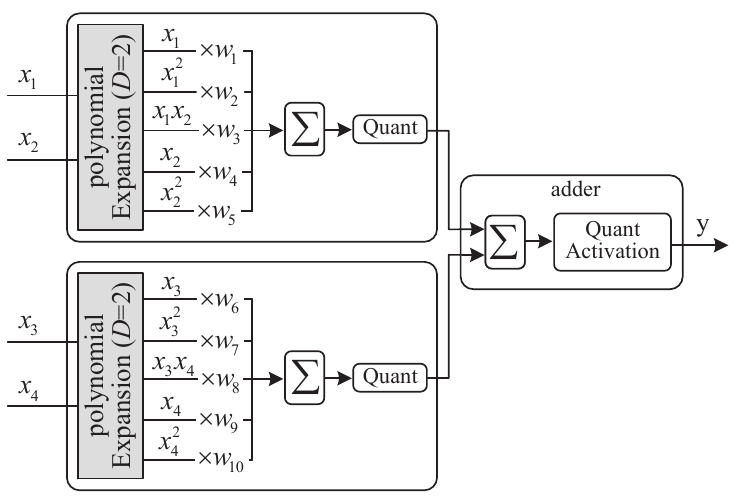}}
\caption{The example of the neuron transfer function.}
\label{fg:lut2}
\end{figure}

Fig.~\ref{fg:lut1} shows the LUT-based MLP approach. In the left part, a maximum of $F$ inputs are selected from ${N}$ nodes of the current layer to connect to each neuron of the next layer (only a single neuron of the next layer is demonstrated here). Additionally, the bit width of each neuron's inputs and outputs is quantized as $\beta$, and the rest of the parameters inside the neuron transfer function are maintained in full precision. Therefore, after training, the transfer function mapping an input vector $\left[ {{x_1},{x_2}, \ldots ,{x_{F}}} \right]$ to the output node can be implemented using $\beta F$ inputs in hardware, and hence its implementation requires $\mathcal{O}(2^{\beta F})$ LUTs in a pre-calculated truth table. 

In practice, an extremely sparse rate is normally applied (size $F \ll N$), which limits the accuracy. Therefore, we introduced the state-of-the-art LUT optimization technique from PolyLUT~\cite{polylut} and PolyLUT-Add~\cite{polylutadd} to compensate for accuracy degradation. As shown in Fig.~\ref{fg:lut2}, the piecewise polynomial functions (polynomial degree is referred to as $D$)~\cite{polylut} and sub-neuron decompose architecture (sub-neuron number is referred to as $A$)~\cite{polylutadd} are used to improve the representation ability and neuron connectivity, respectively. In this example with $F,D,A=2$. The multiplicative combinations of the inputs is expanded from $\left[ {{x_1},{x_2}} \right] \to \left[ {1,{x_1},{x_2},x_1^2,{x_1}{x_2},x_2^2} \right]$ and from  $\left[ {{x_3},{x_4}} \right] \to \left[ {1,{x_3},{x_4},x_3^2,{x_3}{x_4},x_4^2} \right]$. It also generates three decoupled truth tables to reduce the total number of LUTs to $\mathcal{O}(2 \times 2^{\beta F} + 2^{2(\beta +1)})$.

\begin{figure}
\centerline{\includegraphics[width=0.6\linewidth]{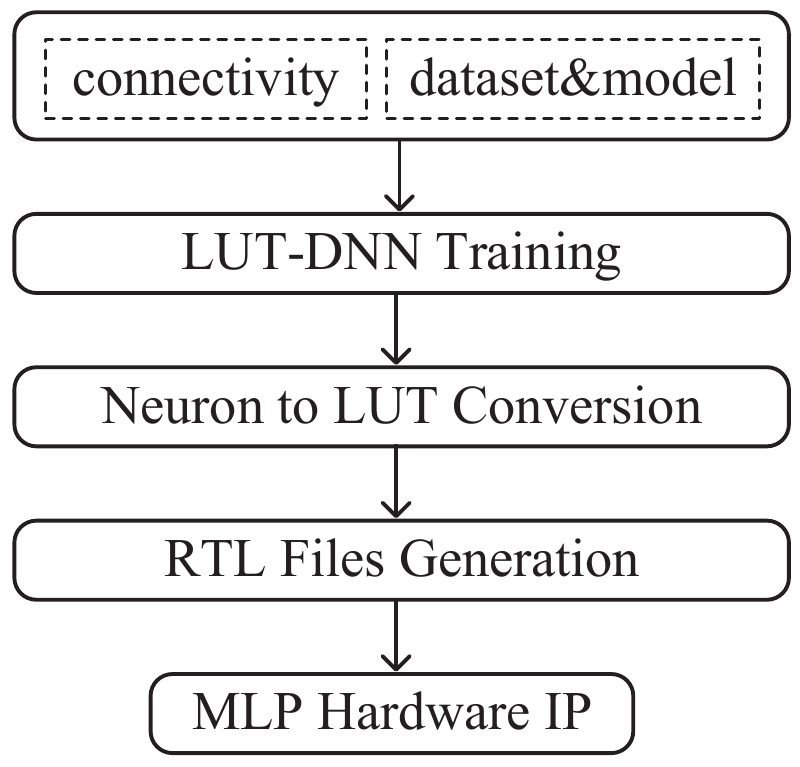}}
\caption{The workflow of the LUT-MLP.}
\label{fg:workflowmlp}
\end{figure}

Fig.~\ref{fg:workflowmlp} shows the tool flow. Given the dataset and model setup, the connectivity (the selected subset in Fig.~\ref{fg:lut1}) can be either randomly selected or optimized in a systematic pruning technique. Training is done offline using PyTorch~\cite{NEURIPS2019_9015}, and the model's weights are transformed into lookup tables following the training phase. Register Transfer Level (RTL) files are then generated in Verilog using these tables, which could be encapsulated into hardware IP using the AMD/Xilinx Vivado tool~\cite{Vivado}.

\subsection{ViT Accelerator}
\label{se:Vision Transformer Accelerator}

Our ViT model builds upon the Transformer architecture from Ref.~\cite{Filip2022tnn}, originally designed for jet-tagging in physics experiments, with key simplifications and optimizations for FPGA deployment. Specifically, (1) the SiLU activation function is replaced with ReLU for more efficient FPGA mapping, and (2) layer normalization is substituted with batch normalization.

Additionally, we adapt this Transformer block for qubit state detection, formulating it as a Vision Transformer, as detailed below.

\begin{figure*}[]
\centerline{\includegraphics[width=1.0\linewidth]{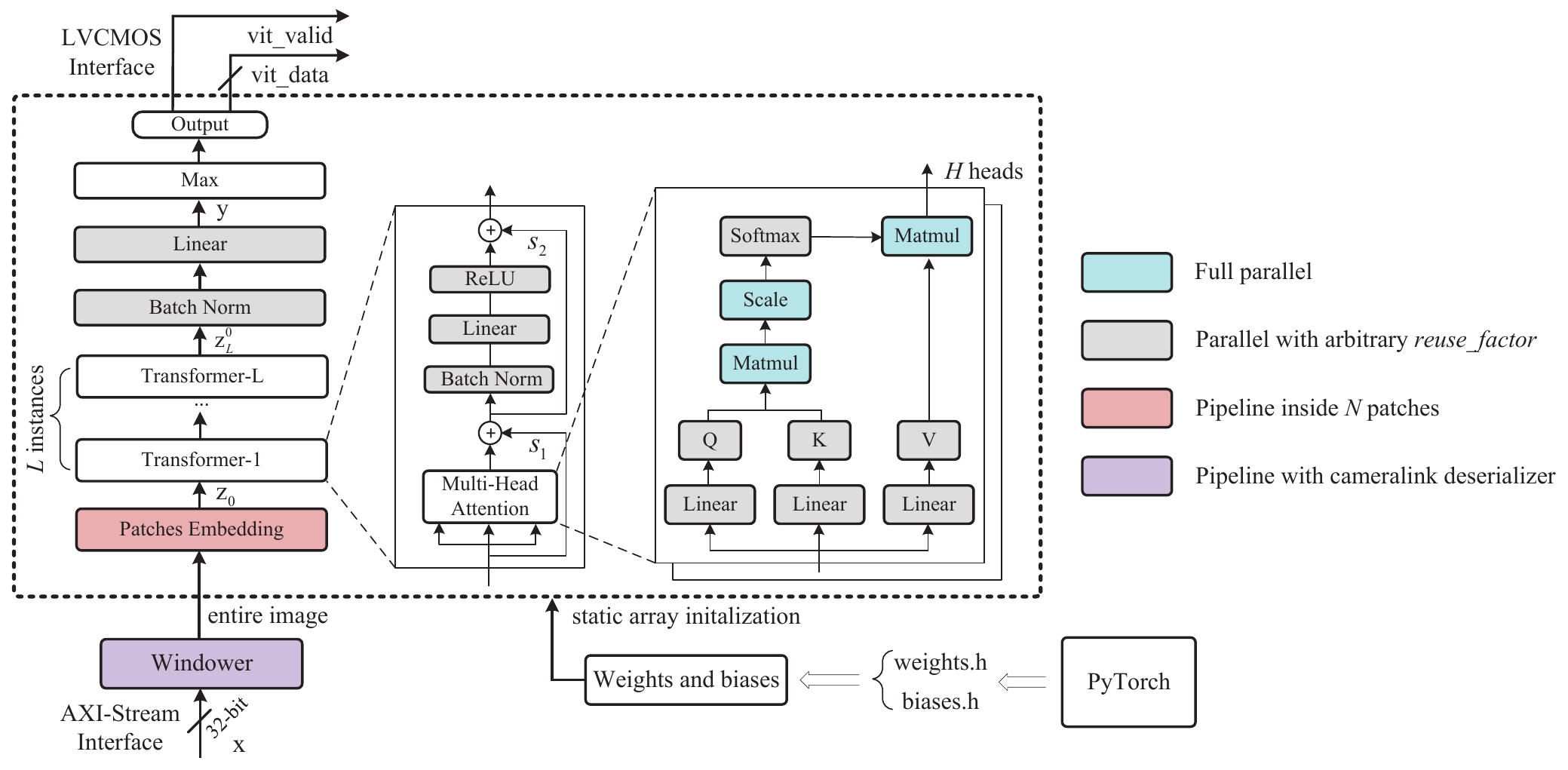}}
\caption{Hardware optimization for each component inside the DNN}
\label{fg:vit_opt}
\end{figure*}

The 2D input image of size ${\bf{x}} \in {\mathbb{R}^{H \times W}}$ is first reshaped into a sequence of flattened 2D patches ${{\bf{x}}_p} \in {\mathbb{R}^{N \times {P^2}}}$, where $\left( {H,W} \right)$ is the resolution of the original image and $\left( {P,P} \right)$ is the resolution of each image patch. $N = HW/{P^2}$ is the resulting number of patches and also serves as the effective input sequence length for the Transformer. As formulated in Eq.~\eqref{eq:z0}, we use ${\bf{x}}_p^n \in {\mathbb{R}^{{P^2}}},n = 1,2,..,N$ to represent $N$ independent patches. The flattened patches are then embedded with an embedding matrix ${\bf{E}} \in {\mathbb{R}^{{P^2} \times D}}$, where $D$ is a constant latent vector size throughout all subsequent Transformer layers. A learnable embedding ${\bf{z}}_0^0 = {{\bf{x}}_{class}}$ is also attached to the sequence of embedded patches in order to integrate global information from the image for the final classification task. For a general ${\bf{z}}_{layer}^{index}$, $layer$ corresponds to the layer index, and $index$ corresponds to the index of the inside component. Finally, the standard learnable 1D position embeddings, ${{\bf{E}}_{pos}} \in {\mathbb{R}^{(N + 1) \times D}}$, are added to the patch embeddings to retain positional information. The resulting sequence of embedding vectors ${{\bf{z}}_0}$ serves as an input to the following Transformer block.

 \begin{eqnarray}  
{{\bf{z}}_0} = \left[ {{{\bf{x}}_{class}};{\bf{x}}_p^1{\bf{E}};{\bf{x}}_p^2{\bf{E}}; \cdots ;{\bf{x}}_p^N{\bf{E}}} \right] + {{\bf{E}}_{pos}}~\label{eq:z0}\\
{{\bf{z}}_l} = {\text{ReLU}}\left( {\text{Linear}}\left( {{\text{BN}}\left(  {\text{MSA}}\left( {{{\bf{z}}_{l - 1}}} \right) + {{{\bf{s}}_{1}}} \right)} \right)  \right)+ {{{\bf{s}}_{2}}}~\label{eq:zl}\\
{\bf{y}} =  {\text{Linear}}\left( {\text{BN}}\left( {{\bf{z}}_L^0} \right) \right)~\label{eq:y}\\
Q_i = {{\bf{z}}_{l-1}}{\bf{W}}_{Qi},\ K_i = {{\bf{z}}_{l-1}}{\bf{W}}_{Ki},\ V_i = {{\bf{z}}_{l-1}}{\bf{W}}_{Vi}~\label{eq:QKV}\\
\text{Attention}(Q, K, V) = \text{softmax}( \frac{QK^T}{\sqrt{d}} )V~\label{eq:attention}
\end{eqnarray}

Each Transformer consists of three major processing elements: Multi-headed Self-Attention (MSA), Batch Normalization (BN), and Linear Layer with ReLU activation function (see Eq.~\eqref{eq:zl}). Also the ${\bf{s}}_{1}$ and ${\bf{s}}_{2}$ provide shortcut connections (${\bf{s}}_{1}$ = ${\bf{z}}_{l - 1}$; ${\bf{s}}_{2}$ = ${\text{MSA}}\left( {{{\bf{z}}_{l - 1}}} \right) + {{{\bf{s}}_{1}}}$). 

Finally, the first element (class token) in the output of the $L$-th transformer layer, ${{\bf{z}}_L^0}$, is processed to produce the final output $\bf{y}$ (see Eq.~\eqref{eq:y}). Categorical labels are therefore calculated by maximizing the output probability (i.e., `dark' or `bright' for each ion).

The MSA is composed of $H$ parallel self-attention (SA) blocks (see Fig.~\ref{fg:vit_opt}). Each SA is then composed of linear layers, matrix multiplication, and a softmax function. The SA is formulated in Eq.~\eqref{eq:QKV} and Eq.~\eqref{eq:attention}, where $Q$, $K$, and $V$ correspond to $queries$, $keys$, and $values$ respectively.
$Q_{i}, K_{i}, V_{i} \in \mathbb{R}^{(N+1) \times D}$ are calculated using matrix multiplications of an input sequence ${\bf{z}} \in {\mathbb{R}^{(N+1) \times D}}$ with weight matrices ${\bf{W}}_{Qi}, {\bf{W}}_{Ki}, {\bf{W}}_{Vi} \in \mathbb{R}^{D \times D}$ (see Eq.~\eqref{eq:QKV}). The Attention function is defined in Eq.~\eqref{eq:attention}, where $d$ is the dimension of the $queries$/$keys$/$values$.

\begin{figure}
\centerline{\includegraphics[width=0.99\linewidth]{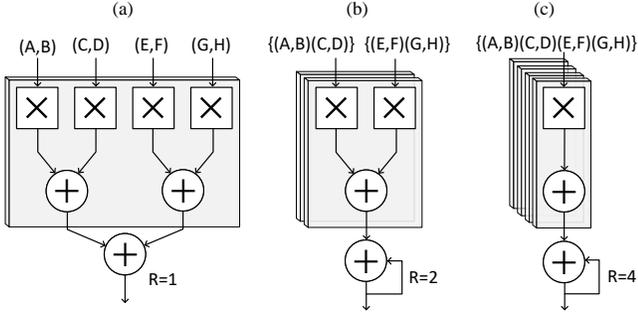}}
\caption{(a). $R=1$, the process is fully parallelized, allowing all multiplication tasks to be carried out at once with the greatest number of multipliers. (b). $R=2$, $\frac{1}{R}$ of the tasks are processed simultaneously, and half of the resources are time-division multiplexed. (c). $R=4$, 4 multipliers are time-division multiplexed, turning the computation into sequential execution.
}
\label{fg:reuse_factor}
\end{figure}

Clearly, ViT's computation is more complex than the transfer function of the MLP; we thus employ a high-level synthesis (HLS) tool to efficiently generate and verify ViT's hardware, rather than directly programming it in RTL. First, we develop a C program-based ViT model that mirrors the PyTorch Python model. Next, the RTL code is generated from the C program using the AMD/Xilinx HLS tool~\cite{VivadoHLS}. Finally, the RTL code is processed through Xilinx Vivado~\cite{Vivado} for hardware IP encapsulation.

We adopted a matrix operation optimization to achieve a favorable resource and latency trade-off on our FPGA platform. It is managed by parameterizing the parallelism degree for the matrix-vector multiplication. Fig.~\ref{fg:reuse_factor} provides an example of artibaried setable \texttt{reuse\_factor} ($R$) with $R$ is set to 1,2,4. In practice, $R$ should be chosen as a number that can be evenly divided~\cite{que2024ll}.

Fig.~\ref{fg:vit_opt} illustrates the ViT accelerator. The first data operation stage is the Windower, which accepts pixel samples from the input AXI-stream interface and uses a shift register to assemble a full image of past inputs to the next stage. Pipeline data transfer is adopted between the Windower and the cameralink deserializer. The Patches Embedding marked in red uses a pipelining scheme over the $N$-sized patches. Modules with high resource consumption and latency, marked in gray, all utilized an arbitrary \texttt{reuse\_factor} to allow their performance to be controlled.

\section{Experimental Setup}
\subsection{Trap ion and EMCCD}

\begin{figure}[]
\centerline{\includegraphics[width=0.53\linewidth]{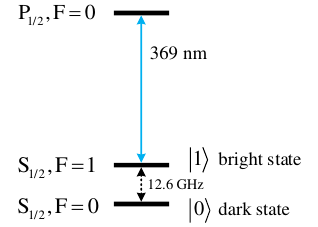}}
\caption{Electronic level structure of ${}^{171}{\text{Y}}{{\text{b}}^ + }$. F indicates the hyperfine quantum number. The 369.5 nm transition is used for Doppler cooling to near the motional ground state and for qubit measurement. The state is measured by detecting photons scattered from the separate lasers on this wavelength. The $^2S_{1/2}$ hyperfine splitting (12.6 GHz) defines two measurable qubit states: bright and dark. Quantum gate operations are performed using an off-resonant 355 nm optical transition (not shown).}
\label{fg:171Yb}
\end{figure}

The system is designed for a macroscopic radio-frequency Paul trap, using ${}^{171}{\text{Y}}{{\text{b}}^ + }$ as the trapped ion set. The $\ket{0}$ and $\ket{1}$ qubit states are encoded in the $\ket{^2\text{S}_{1/2},\text{F}=0}$ and $\ket{^2\text{S}_{1/2},\text{F}=1}$ energy levels of the valence electron, respectively (see Fig.~\ref{fg:171Yb}).

In this work, an Andor 987 EMCCD camera~\cite{andor987} is employed to collect the fluorescence from ${}^{171}{\text{Y}}{{\text{b}}^ + }$ ions and then converts the photons into a digital signal. The EMCCD camera is controlled by the ARTIQ system with a Kasli Controller v1.1~\cite{artiq, Kasli}.

\subsection{Image dataset}

We collect 6,500 single-ion images from the Paul trap setup and label them to construct a 1-qubit dataset. To ensure a balanced distribution, approximately 50\% of the images are labeled as `bright'. Each image has a resolution of \(10 \times 10\) pixels and is divided into training and testing datasets with a 9:1 ratio.  

\begin{figure}
\centerline{\includegraphics[width=10 cm]{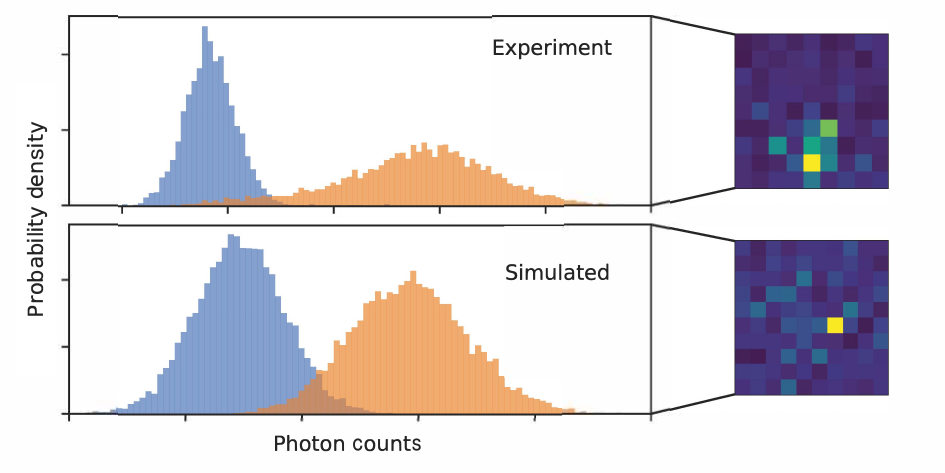}}
\caption{Insets show an example of a synthetic and experimental image of a single ion.}
\label{fg:sim_vs_exp_hists}
\end{figure}

For multi-ion scenarios, the trapped ion system used in our setup only allows for pre-set labels that are globally applied to the entire ion chain. For instance, in a 3-ion configuration, only the labels `000' and `111' can be assigned before the experiment. Consequently, we lack access to the real label distribution in experimentally collected multi-ion images. To facilitate DNN training and latency evaluation, we generate synthetic multi-ion images in three steps. First, the ideal ion fluorescence is modeled as a Gaussian distribution with a standard deviation of \(\sigma_x = \sigma_y = 0.4\) pixels and an amplitude of 35. Second, multiplicative Poisson noise is added, where the average noise intensity is modeled using \texttt{numpy.random.poisson}(\(\lambda = 0.5\)). Finally, Gaussian background noise with a mean of \(\mu_{bg} = 50\) and a standard deviation of \(\sigma_{bg} = 20\) is introduced across the entire image. As illustrated in Fig.~\ref{fg:sim_vs_exp_hists}, these parameters are selected to ensure that the histogram statistics of the synthetic data align with experimentally calibrated values. Additionally, a spatial analysis of experimental images confirms that ion fluorescence can be well approximated by a Gaussian distribution.  

The resulting 3-qubit dataset consists of 100,000 images, each with a resolution of \(12 \times 24\) pixels. The three ions are spaced 5 pixels apart, consistent with typical ion separations observed in experimental setups.

\subsection{Latency Measurement Setup}

\begin{figure}[]
\centerline{\includegraphics[width=1.0\linewidth]{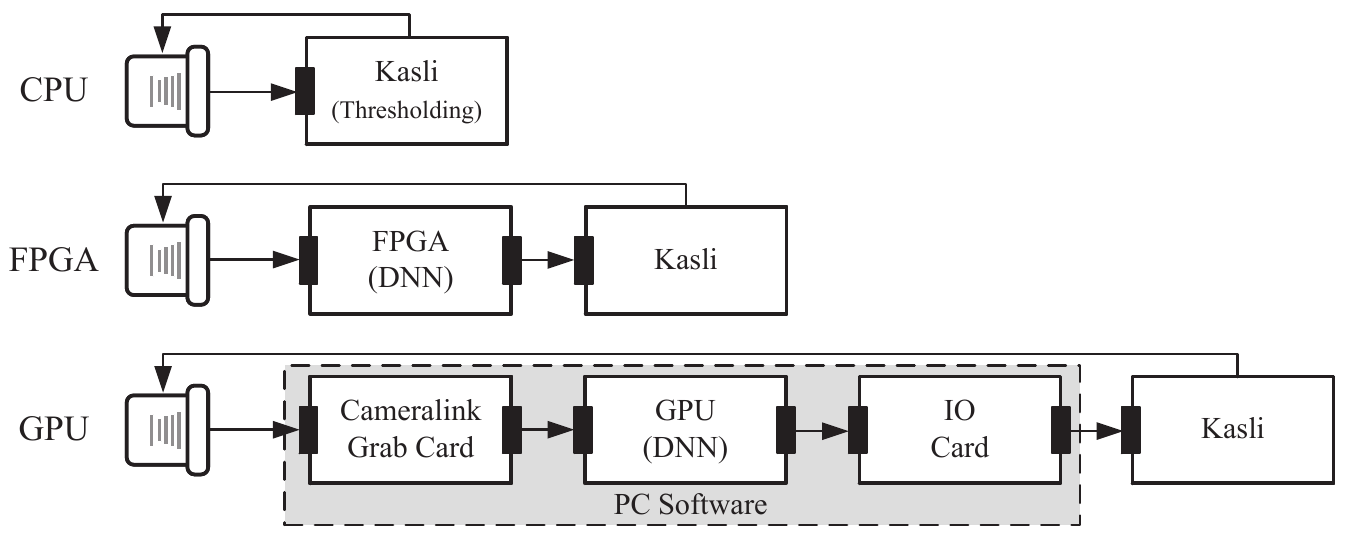}}
\caption{This figure shows three different experiment setups. (a) is the Thresholding method implemented on Kasli. (b) is our FPGA solution with DNN as the classifier. (c) is the GPU solution; The detailed dataflow of (b) and (c) is shown in Fig.~\ref{fg:gpu}.}
\label{fg:3setup}
\end{figure}

We compare latency performance using three different setups (see Fig.~\ref{fg:3setup}):\\
(a). Thresholding implemented on Kasli (CPU).\\
(b). DNN implemented on FPGA.\\
(c). DNN implemented on GPU.

In the baseline system, Kasli deploys thresholding as an integral qubit detection method. An example of thresholding is illustrated in Fig.~\ref{fg:sim_vs_exp_hists}. In the calibration process, we process histograms for the bright (orange) and dark images (blue) and determine an optimal threshold that maximizes the fidelity~\cite{halama2022real}. Afterward, a measurement is processed by calculating the total number of photon counts in the image and comparing it to the pre-calculated threshold. 

In contrast, the proposed FPGA qubit detection system uses the Sundance SOLAR EXPRESS 120 (SE120), containing an Xilinx XCZU11EG chip on which the DNN accelerator is implemented. Images from the EMCCD camera are captured using a Cameralink FPGA Mezzanine Card Module~\cite{sundance}. 

As for GPU solutions, given that commercial GPU products do not have a cameralink interface and user-programable IO interface. To allow the GPU to process image data from the EMCCD and send out classification results to the Kasli controller in the lowest latency, we assembled the 3-card solution in Fig.~\ref{fg:3setup} (c).

\begin{figure}[]
\centerline{\includegraphics[width=1.0\linewidth]{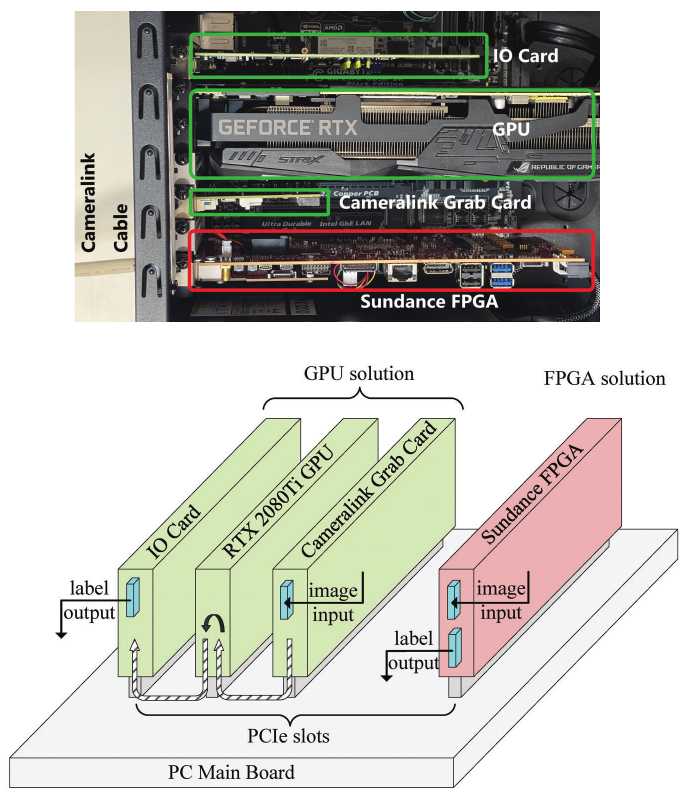}}
\caption{The dashed arrow indicates the data flow inside the GPU solution. The FPGA is only powered and controlled by the PC Mainboard; no data transmission is issued between them. }
\label{fg:gpu}
\end{figure}

Fig.~\ref{fg:gpu} demonstrates the board's setups and datapath in our GPU and FPGA experiments. The GPU and FPGA tests are on the same PC mainboard for consistent evaluation; the Cameralink Cable is either connected to the GPU or FPGA setups for independent tests. We use a commercial off-the-shelf (COTS) card to receive cameralink images, PyTorch with GPU for ML inference, and a COTS GPIO card for label outputs. Teledyne Dalsa Aquarius Base CL x1 is used as the cameralink grab card~\cite{Teledyne}, an NVIDIA GeForce RTX 2080 Ti GPU is located in the middle, followed by the Advantech PCIE-1751-B card as the GPIO card~\cite{Advantech}. All three cards equip the PCIe point-to-point host interface, which allows simultaneous image acquisition, ML inference, and label output with little intervention from the host CPU. These three cards are mounted in a host PC main board with Intel(R) Core(TM) i7-4790 @3.6GHz and 16GB memory.

\subsection{DNN Model configuration}
The MLP classifier is configured with 5 layers, and the number of neurons in each layer is 256,100,100,100,10. The polynomial degree $D=2$ and sub-neuron adder $A=2$ are applied. The learning rate of $lr=0.008$ with the AdamW optimizer is used in training. The $\beta=2$ and $F=4$ are applied for efficient computation.

The size of the ViT classifier model is ($L=1$, $H=8$, $D=16$), where $L$ is the number of transformer layers, $H$ is the number of heads, and $D$ is the latent dimensions of self-attention. The patch size is set to $P=5$ for 1-qubit classification and $P=6$ for 3-qubit classification. We use a batch size of 128 and a learning rate $lr=0.05$ for the SGD optimiser~\cite{NEURIPS2019_9015}. The ViT is trained and tested using Pytorch~\cite{NEURIPS2019_9015}. 

The chosen hyperparameters offer good detection performance and efficient hardware implementation. Further optimization may improve the solution for specific system requirements. The ViT parameters were represented in Vivado HLS using a fixed point format \texttt{ap\_fixed$\langle 16,8\rangle$} with a 16-bit word length and an 8-bit fraction field.

\section{Results}
\subsection{Fidelity Comparison}

In this section, we evaluate the fidelity achieved by our DNN model for 1- and 3-ion images. On ARTIQ, the region of interest (ROI) in Thresholding is set to (width=4, height=8) around the center point of each ion.

\begin{equation}
    \label{eq:fidelity}
\bar F = \frac{1}{{{2^n}}}\sum\limits_i {p\left( {measured{\text{ }}i|prepared{\text{ }}i} \right)}
\end{equation}

The mean measurement fidelity (MMF) is employed as a fidelity comparison, as defined in Ref.~\cite{seif2018machine}, which is formulated as Eq.~\eqref{eq:fidelity} where $n$ is the number of ions, and the summation index $i$ spans across all possible combinations of $n$-qubit states. The measured $i$ and prepared $i$ represent predicted output from the qubit classifier and the provided label, respectively. The classification error is therefore calculated as (1 - $\bar F$)

\begin{table}[]
\centering
  \caption{Comparison of three detection models in accuracy and parameters}
  \scalebox{1.0}{
  \renewcommand{\arraystretch}{1.005}
  \setlength{\tabcolsep}{0.8cm}
  \resizebox{0.98\columnwidth}{!}{ 
  \begin{tabular}{|c|c|c|c|}
  \hline
  \multicolumn{1}{|c|}{\textbf{Dataset}}    & \textbf{Model} &  \textbf{MMF Error} (\%)$\downarrow$ \\ \hline \hline
  \multirow{3}{*}{\makecell[c]{1-Qubit\\($10 \times 10$)}}  & Threshold & 2.0    \\ 
    &  MLP  & 1.1     \\
    &  ViT  & 0.8       \\ \hline
    \multirow{3}{*}{\makecell[c]{3-Qubit\\($12 \times 24$)}}  & Threshold & 11.4     \\ 
    &  MLP  & 2.7     \\
    &  ViT  & 1.5       \\ \hline
\end{tabular}}}
\label{tb:accuracy}
\end{table}

The MMF error for the 1-qubit and 3-qubit datasets obtained with thresholding, MLP, and ViT is reported in Table~\ref{tb:accuracy}. 
For a 1-qubit, the MMF error of MLP and ViT is lower than the thresholding error by 0.9\% and 1.2\%, respectively. In the 3-qubit scenario, the benefits of the DNN models become more apparent, as MLP and ViT yield MMF errors that lower the threshold method by 8.7\% and 9.9\%. The gap between the two DNN models is due to their complexity.

\subsection{Latency Comparison}
\label{se:4_latency_comparison}

The FPGA design is running under a frequency constraint of 250 MHz (4 $ns$); the resource consumption of synthesized hardware is shown in Table~\ref{tb:resource_4ns}. MLP consumes 5 clock cycles (20 $ns$) for 1- and 3-qubit cases, while ViT, implemented under 1-qubit and 3-qubit sets, consumes 4054 (16.22 $\mu s$) and 8797 clock cycles (35.19 $\mu s$), respectively.

\begin{table}[]
\centering
  \caption{LUT, Digital Signal Processing (DSP) Elements, Block RAM (BRAM), and Flip Flops (FF) are used to indicate the main FPGA programmable logic resources.}
  \scalebox{1.0}{
  \renewcommand{\arraystretch}{1.005}
  \setlength{\tabcolsep}{0.35cm}
  \resizebox{0.98\columnwidth}{!}{ 
  \begin{tabular}{|l|c|c|c|c|}
  \hline
  \textbf{Mode}    & \textbf{LUT} &  \textbf{DSP} & \textbf{BRAM} & \textbf{FF} \\ \hline \hline
  Cameralink  & 1513 & 0 & 18 & 2097    \\ 
  \textbf{Util}  & 0.5\% & 0\% & 1.5\% & 0.35\%   \\ \hline
  MLP (1-Qubit)  & 15500 & 0 & 0 & 3000    \\ 
  \textbf{Util}  & 5\% & 0\% & 0\% & 0.5\%   \\\hline
  ViT (1-Qubit)  & 69395 & 467 & 207 & 48113    \\ 
  \textbf{Util}  & 23\% & 15\% & 17\% & 8\%   \\ \hline
  MLP (3-Qubit)  & 16065 & 0 & 0 & 3150    \\ 
  \textbf{Util}  & 6\% & 0\% & 0\% & 0.5\%   \\\hline
  ViT (3-Qubit)  & 152021 & 600 & 272 & 76568    \\ 
  \textbf{Util}  & 50\% & 20\% & 22\% & 12\%   \\ \hline
\end{tabular}}}
\label{tb:resource_4ns}
\end{table}

System latency is experimentally determined by measuring the duration between various probed signals on an external oscilloscope (a 2.5 GHz oscilloscope for dual-channel edge detection). The exposure time of the EMCCD camera is set to 1 $ms$, and the Cameralink transmission frequency is set to its maximum frequency of 17 MHz~\cite{andor987}. The trigger signal from ARTIQ to start the camera acts as the starting point for latency measurements.

\begin{itemize}
\item \texttt{FVAL}  : ${{\text{T}}_{{\text{Ch2}}}}$(\texttt{FVAL}) $-$ ${{\text{T}}_{{\text{Ch1}}}}$(\texttt{trigger})

\item \texttt{tx\_done} : ${{\text{T}}_{{\text{Ch2}}}}$(\texttt{tx\_done}) $-$ ${{\text{T}}_{{\text{Ch1}}}}$(\texttt{trigger})

\item \texttt{vit\_valid}: ${{\text{T}}_{{\text{Ch2}}}}$(\texttt{vit\_valid}) $-$ ${{\text{T}}_{{\text{Ch1}}}}$(\texttt{trigger})
\end{itemize}

Since the MLP incurs only nanosecond-level latency, we use the ViT model to benchmark the latency of FPGA- and GPU-based implementations, as shown in Table~\ref{tb:vit-latency}. The \texttt{FVAL} signal, measured from the FPGA's probed I/O, is 1.41$\pm$0.01 $ms$. As this latency is dictated solely by the Cameralink protocol, we assume the GPU solution shares the same value, given that direct measurement is not feasible due to the lack of accessible probed I/O on the GPU.

The \texttt{tx\_done} measured on FPGA is conducted on target image sizes $10 \times 10$ and $12 \times 24$, which are 1.75$\pm$0.01 $ms$ and 2.25$\pm$0.02 $ms$ respectively. Unfortunately, the precise value for the GPU solution is also immeasurable.

The overall system latency indicated by \texttt{DNN\_valid} shows that the FPGA system achieves $119\times$ and $94\times$ speedup over the GPU solution for 1-qubit and 3-qubit tests, respectively. 

Interestingly, the ML accelerator on the GPU accounts for only $\sim$1.3\% of the total qubit detection duration, while $\sim$98\% of the latency is dominated by the PCIe-rooted interface. This explains why the latency of the GPU-based solution is indeterminate and exhibits significant variation.

\begin{table}[]
\centering
  \caption{Latency measurement results of FPGA- and GPU-based qubit detection system (Unit: $ms$, model: ViT). The mean and standard deviation are calculated from 20 repeated tests. The FPGA-based ViT latency (0.016 $ms$ and 0.035 $ms$) listed here is from the Vivado hardware report, which consumes fixed clock cycles (4054 for 1-qubit and 8797 for 3-qubit) in repeated executions.}
  \scalebox{1.0}{
  \renewcommand{\arraystretch}{1.105}
  \setlength{\tabcolsep}{0.1cm}
  \resizebox{0.98\columnwidth}{!}{ 
  \begin{tabular}{|l|c|c|c|c|}
  \hline
  \textbf{Platform}  & \multicolumn{2}{c|}{FPGA} & \multicolumn{2}{c|}{GPU} \\ \hline
  Ion number  & 1-Qubit & 3-Qubit & 1-Qubit & 3-Qubit  \\ \hline  \hline
  \texttt{FVAL}  & 1.41$\pm$0.01  & 1.41$\pm$0.01  & 1.41$\pm$0.01     & 1.41$\pm$0.01   \\
 \texttt{tx\_done}  & 1.76$\pm$0.01  & 2.25$\pm$0.02  & -                 & -    \\ 
  \texttt{DNN\_valid} & 1.78$\pm$0.02  & 2.29$\pm$0.02  & 211.95$\pm$17.72  & 214.70$\pm$10.87   \\ \hline
  ViT model  & 0.016 & 0.035 & 2.85$\pm$0.13 & 2.95$\pm$0.11    \\ 
  \textbf{Util}  & 0.89\% & 1.53\% & 1.34\% & 1.37\%   \\ \hline
\end{tabular}}}
\label{tb:vit-latency}
\end{table}

\begin{figure}[h]
\centering
\includegraphics[width=1.0\linewidth]{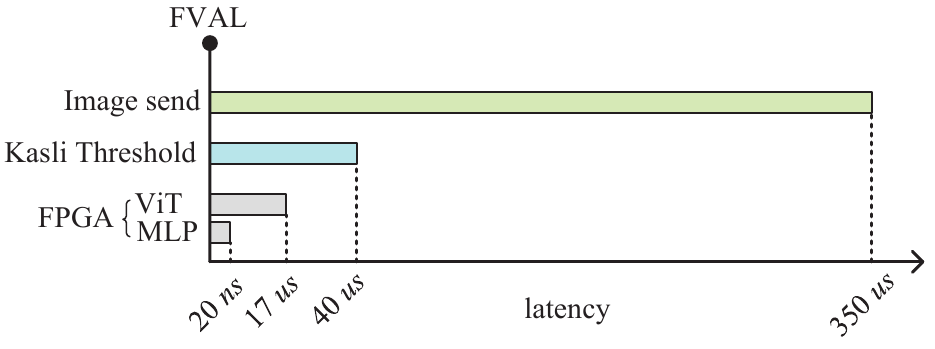}
\caption{Latency Measurements from FVAL for 1-qubit Detection.}
\label{fg:system_latency}
\end{figure}

\begin{figure*}[h]
\centerline{\includegraphics[width=15cm]{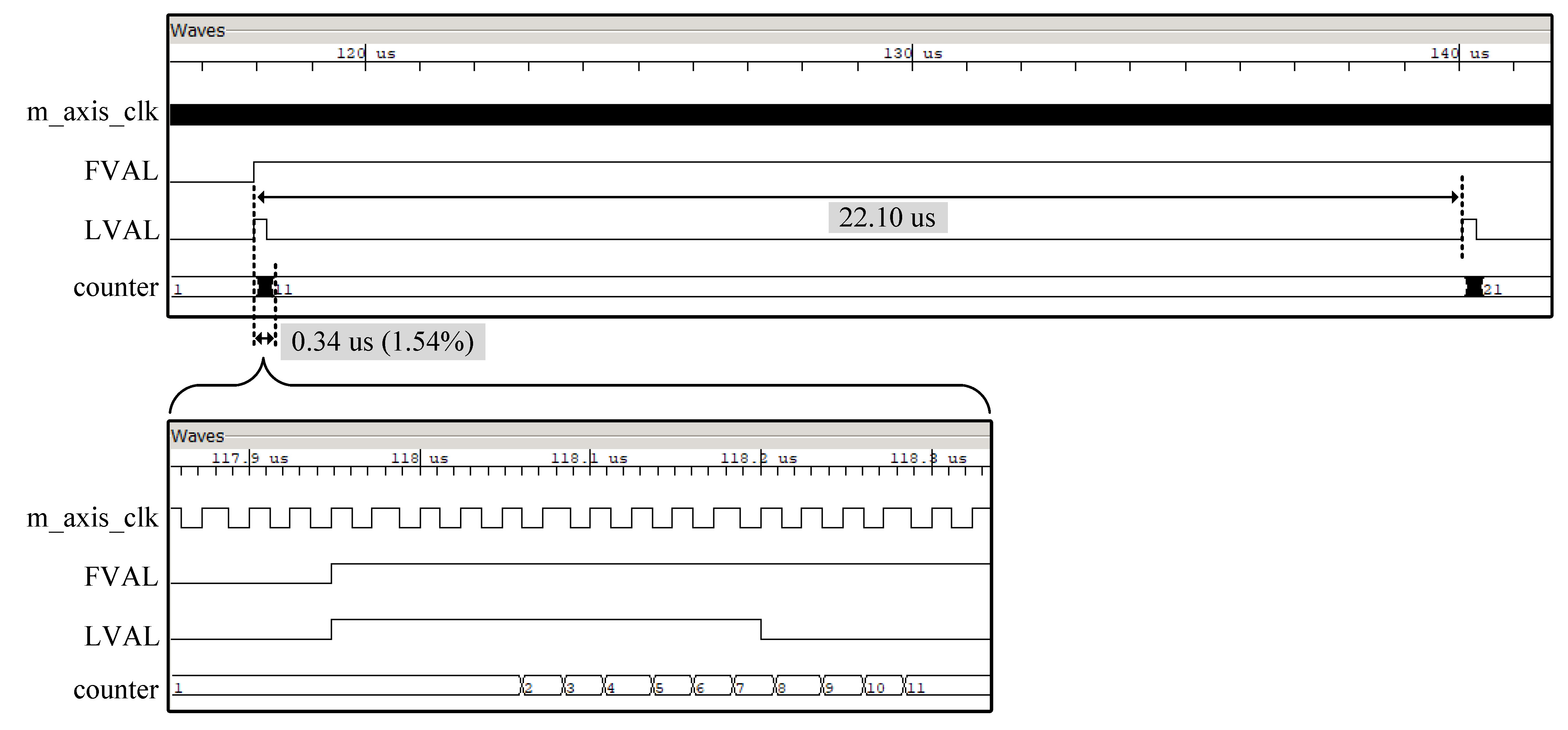}}
\caption{ILA Waveform over an entire $10 \times 10$ pixel image capture. It is captured in real-time at 250 MHz and reflects the actual data transmission. We only show a single line duration (\texttt{LVAL} is valid) for clear illustration purposes. The counter indicates the number of received pixels inside the DNN. It accepts a new row of 10 pixels each time \texttt{LVAL} is valid.}
\label{fg:ila_waveform}
\end{figure*}

Although the FPGA-based solution achieves a significant speed-up compared to the GPU, the dominant source of latency lies in data movement rather than DNN inference. To address this, Fig.~\ref{fg:system_latency} presents a breakdown of transmission and computation latency on the FPGA, using a single-qubit example. 

The image send latency is measured between \texttt{tx\_done} and \texttt{FVAL}, which is determined by the camera itself. To further analyze this, we used the integrated logic analyzer (ILA~\cite{VivadoILA}) to capture internal signals during the image transmission period. As shown in Fig.~\ref{fg:ila_waveform}, the actual line reading operation accounts for only 1.54\% of the interval between two \texttt{LVAL} signals, while dummy transmissions entirely occupy the remaining time.

This inefficiency arises because each line requires 512 pixel-time slots~\cite{andor987}, even though only 10 pixels are transmitted. If this limitation were addressed, the image send could be completed in just 5.45 $\mu s$---a 65$\times$ speedup. This observation highlights a key direction for future improvements: optimizing camera readout could significantly reduce latency in CCD-based qubit measurement systems.

\section{Conclusion}

This paper presents a neural network-based qubit detection system and benchmarks its real-time latency performance for CCD-based trapped-ion quantum information processing on both FPGA and GPU platforms.

On the FPGA, we first reduce data buffer latency by implementing a direct interface between the EMCCD camera and the FPGA. We employ DNNs (MLP and ViT) to infer qubit states from 2D images, both of which demonstrate improved fidelity compared to conventional thresholding~\cite{halama2022real}. The LUT-based MLP achieves an ultra-low latency of 20 $ns$, significantly lower than the 40 $\mu s$ latencies measured in the Thresholding baseline. Meanwhile, the ViT model offers a more balanced trade-off between latency (16-35 $\mu s$) and accuracy.

Experimental results show that the total detection latency on the FPGA is two orders of magnitude lower than that of the GPU. This improvement stems from the FPGA’s highly parallel architecture, elimination of multi-card data transfers, and minimal buffering overhead.

Interestingly, in the proposed FPGA system, image transmission latency dominates the overall detection pipeline. The bottleneck lies in the EMCCD camera’s image digitization and transmission via Cameralink. This observation suggests that achieving further latency reductions will necessitate advancements in camera readout interfaces in future work. Moreover, while the current implementation is optimized for the available hardware, the proposed architecture is not limited to a specific FPGA design. Future qubit detection systems with faster camera interfaces and improved FPGA platforms can adopt the same architecture, benefiting from larger, higher-performance FPGAs to further reduce latency and enhance scalability.

\newpage
\bibliographystyle{unsrt}
\bibliography{ref}

\end{document}